\RequirePackage{amsthm}
 
\documentclass[referee, pdflatex,sn-mathphys-num]{sn-jnl}

\usepackage[T1]{fontenc}
\usepackage[utf8]{inputenc}

\usepackage{graphicx}%
\usepackage{multirow}%
\usepackage{amsmath,amssymb,amsfonts}%
\usepackage{amsthm}%
\usepackage{mathrsfs}%
\usepackage[title]{appendix}%
\usepackage{xcolor}%
\usepackage{textcomp}%
\usepackage{manyfoot}%
\usepackage{booktabs}%
\usepackage{algorithm}%
\usepackage{algorithmicx}%
\usepackage{algpseudocode}%
\usepackage{listings}%
\usepackage{siunitx}

\theoremstyle{thmstyleone}%
\theoremstyle{thmstyletwo}%

\theoremstyle{thmstylethree}%

\begin{document}

\title[Physics-based Modeling and Simulation of Nanoparticle Networks]{Physics-based Modeling and Simulation of Nanoparticle Networks}


\author*[1]{\fnm{Torben} \sur{Hemke}}\email{torben.hemke@ruhr-uni-bochum.de}

\author[1]{\fnm{Robin} \sur{Struck}}\email{robin.struck@ruhr-uni-bochum.de}

\author[1]{\fnm{Sahitya} \sur{Yarragolla}}\email{sahitya.yarragolla@ruhr-uni-bochum.de}

\author[2]{\fnm{Tobias} \sur{Gergs}}\email{tog@tf.uni-kiel.de}

\author[2,3]{\fnm{Jan} \sur{Trieschmann}}\email{jt@tf.uni-kiel.de}

\author*[1]{\fnm{Thomas} \sur{Mussenbrock}}\email{thomas.mussenbrock@ruhr-uni-bochum.de}

\affil*[1]{\orgdiv{Chair of Applied Electrodynamics and Plasma Technology}, \orgname{Ruhr University Bochum}, \orgaddress{\street{Universitätsstraße 150}, \city{Bochum}, \postcode{44801}, \state{NRW}, \country{Germany}}}

\affil[2]{\orgdiv{Theoretical Electrical Engineering, Faculty of Engineering}, \orgname{Kiel University}, \orgaddress{\street{Kaiserstraße 2}, \city{Kiel}, \postcode{24143}, \state{Schleswig-Holstein}, \country{Germany}}}

\affil[3]{\orgdiv{Kiel Nano, Surface and Interface Science KiNSIS}, \orgname{Kiel University}, \orgaddress{\street{Christian-Albrechts-Platz 4}, \city{Kiel}, \postcode{24118}, \state{Schleswig-Holstein}, \country{Germany}}}


\abstract{This study presents the computational modeling and simulation of silver nanoparticle networks (NPNs), which, in the realm of neuromorphic computation, suggest to be a promising candidate for nontraditional computation methods. The modeling of the networks construction, its electrical properties and model parameters are derived from well-established physical principles.}




\maketitle

\section{Introduction}\label{sec:introduction}

In the burgeoning field of neuromorphic computing, memristors have emerged as pivotal components due to their ability to emulate the synaptic functions of biological neurons \cite{song_2023, christensen_2022}. These devices, characterized by their memory resistance, are capable of retaining a history of the electrical charge that has passed through them, thus enabling them to mimic the plasticity of synapses\textemdash key to learning and memory in biological systems \cite{chen_2023, nawrocki_2016}. The unique properties of memristors make them highly suitable for neuromorphic applications, where the goal is to replicate the neural structures and computational strategies of the (human) brain in hardware form \cite{park_2022}.
Nanoparticle (NP) \cite{mallinson_2023, bose_2022, carstens_2022, dunham_2021} and nanowire (NW) networks \cite{milano_2024, loeffler_2020, diaz-alvarez_2019} have shown great promise in advancing neuromorphic computing . These networks can be engineered to be at the brink of electrical percolation, a critical threshold where the system transitions from being an insulator to a conductor \cite{carstens_2022}. At this juncture, the network exhibits brain-like critical dynamics and long-range temporal correlations (LRTC), which are closely connected to the computational capabilities and complex, self-organized behavior observed in biological neural networks \cite{chakraborty_2020, ielmini_2020}, suggesting that such systems can process information in a highly efficient and parallel manner, much like the human brain.

The memristive properties of these networks are governed by electrochemical metallization, a process that forms and dissolves conductive filaments, akin to the strengthening and weakening of synapses during learning processes.
The study of percolating nanoparticle networks (NPNs) is particularly intriguing as it offers a pathway to emulate the emergent functional behavior known from biological neuronal networks. 

Applications of memristive NPNs in neuromorphic computing are diverse, ranging from pattern recognition \cite{ielmini_2020} over true random number generators \cite{acharya_2021} to dynamic information processing \cite{heywood_2022}. They offer a distinct class of neuromorphic hardware architectures that provide an alternative to highly regular arrays of memristors. By leveraging the intrinsic properties of these networks, researchers aim to develop systems that not only process information with high efficiency but also adapt and learn from their environment, paving the way for intelligent, autonomous machines.

In conclusion, the exploration of memristive properties in NPNs represents a significant step towards the realization of neuromorphic systems that can operate at the edge of criticality, offering a hardware platform that is both energy-efficient and capable of complex computations. The ongoing research in this domain holds the potential to revolutionize the way we approach computing, leading to devices that operate more like the human brain than traditional transistor-based computers.

In this work, we focus on the computational modeling and simulation of silver (Ag) NPNs, described in \citep{carstens_2021, gronenberg_2024}. Network creation is grounded on gas aggregation source deposition \cite{drewes_2022} via a constant area rate. The NP and their electrical interaction form the core of our simulation. We are expanding current research \cite{pike_2020} with physically based approaches, such as a polydisperse NP diameter distribution, more in-depth models to describe tunneling and filament conductance, and a simple integration of material properties, allowing for easy study of NPNs composed of various substances.

The paper is organized as follows: In the next section we present the model used in this study. The following sections describe the framework conditions of the simulation and interpret the results. The last section summarizes the key findings and provides an outlook on possible subsequent work.

\section{Nanoparticl Network Model}\label{sec:model}
The model used here is applied in \cite{gronenberg_2024} and briefly described there. In this section, we will present a more detailed description. The basic idea is a 2D continuum percolation model of overlapping disks \cite{Mertens2012}. These disks with centers $x_i$, $y_i$ and radii $r_i$ represent the NPs. Typically, the deposited NPs in the experiment follow a specific radius distribution. Our model is capable of accommodating these polydisperse NPs.

\subsection{Modeling NPs deposition as a Poisson process}
First, we define a 2D substrate consisting of SiO\textsubscript{2} of width $w$ in $x-$direction and height $h$ in $y-$direction. We place NPs with conductivity $\sigma$\textsubscript{Ag} stochastically on it. Two assumptions are made: Firstly, an intensity rate $\lambda$ that does not change locally, and secondly, the condition that NPs cannot occur more than once in exactly the same place. This leads to the following Poisson process.
Consider an area $A$ of the substrate and partition it into $n$ subregions. When $n$ goes to infinity the probability of a NP being in one of the subregions is \cite{Gelfand2010}
\begin{equation}
p = \frac{\lambda A}{n}.
\end{equation}
To calculate the probability of $k$ NPs being in $A$ the limit of $n \to \infty$ is taken, see Appendix \ref{secA1} for details,
\begin{eqnarray}
    P_{A}(k) &=& \lim_{n \to \infty} \binom{n}{k} p^k (1-p)^{n-k} \\
    &=& \frac{(\lambda A)^k e^{-\lambda A}}{k!}.
\end{eqnarray}

For the simulation the substrate is partitioned into $N$ cells $A_{i},i=1,2,...N$ that are assigned a number of NPs to contain in accordance to (3). Inside the cells the NPs are distributed uniformly which is an error since this does not match the Poisson process that is also at work inside of the cells. This error gets smaller the bigger $N$ gets as the probability of more than one NP being in one cell decreases rapidly. Thus the error resulting from the NPs being equally distributed inside of the cells decreases as well. To get a concrete result for the cell size the probability of more than one NP being spawned inside of a cell is chosen to be less or equal than 10\%,
\begin{eqnarray}
    0.9 &\leq& P_{A_{i}}(0) + P_{A_{i}}(1) 
    = e^{-\lambda A_{i}} + (\lambda A_{i}) e^{-\lambda A_{i}}.
\end{eqnarray}
This results in the following condition for $\lambda A_{i}$:
\begin{equation}
    \lambda A_{i} \leq 0.5
\end{equation}
The relationship between the NPs coverage $\Phi$ over the substrate can be derived with $\bar{A_{p}}$ being the mean particle area, $A_{s} = wh$ the substrate area, and $N_{p}$ the number of particles by \cite{Mertens2012}:
\begin{equation}\label{eq:coverage}
    \Phi = 1 - e^{- N_p \frac{\bar{A_{p}}}{A_s}}
\end{equation}
This leads to
\begin{eqnarray}
    N_{p} &=& - \frac{A_{s}}{\bar{A_{p}}} \ln \left(1- \Phi \right) \\
    \lambda &=& \frac{N_{p}}{A_{s}} \\
    \Rightarrow \lambda &=& - \frac{\ln \left(1- \Phi \right)}{\bar{A_{p}}}.
\end{eqnarray}
Finally, a result for the cell area $A_{i}$ can be expressed as
\begin{eqnarray}
    A_{i} &\leq& -0.14855 \hspace{0.1cm} \frac{\bar{A_{p}}}{\ln \left(1- \Phi \right)}.
\end{eqnarray}

In addition, we apply another condition for the positioning of the NPs on the substrate: The Ag NPs are most likely to deposit as truncated octahedra \cite{Ngandjong2016}. Thus, they must have at least the distance of the corresponding lattice constant, which is $\approx~0.405$ nm \cite{Chen2017}. NPs below this distance are removed in the deposition algorithm.

\subsection{Grouping and Connections}
After the NPs are deposited as 2D disks following a Poisson distribution across the cells and evenly distributed within them, we iterate through these cells to identify overlapping disks. These overlapping disks are then combined into groups, Fig.~\ref{fig:particles}.
Next, a Delaunay triangulation is performed between the centers of the NPs, and the resulting connections are filtered based on specific criteria, such as the maximum allowed connection length. These connections are assigned electrical conductances, depending on whether the connection is within a group, Sec. \ref{sec:bulk_conductance}, or in a gap, Sec. \ref{sec:gap_switching}.

\subsubsection{Bulk Conductance}\label{sec:bulk_conductance}
We model the conductance of the (Ag) bulk connections using the standard form:
\begin{equation}
G = \sigma \frac{A_\textsubscript{eff}}{d\textsubscript{NP}}
\end{equation}
Here, we calculate the intersection line of the 2D disks representing the NPs and take this as the diameter for an intersection circle of the area $A$\textsubscript{eff}. The length $d$\textsubscript{NP} is approximated by the distance centers of the 2D disks.

\subsection{Gap Switiching Model}\label{sec:gap_switching}
n the initial state, the gap is in tunnel mode and has the corresponding conductance, Sec. \ref{sec:tunnel_conductance}. When the electric field in the gap exceeds the threshold value $E\textsubscript{th}$ the gap begins to close through the growth of a filament. In each iteration, the tunneling conductance is adjusted to the current length of the gap and recalculated. Once the gap is closed, meaning the filament connects both NPs, the corresponding conductance is assigned to the filament (with a parameterizable inital filament diameter), Sec. \ref{sec:filament_conductance}. Due to electromigration and surface diffusion \cite{wu_2022}, the filament can break again, and the tunneling conductance is then reassigned.

\subsubsection{Filament Conductance}\label{sec:filament_conductance}
The conductance of a conductive filament can be defined as follows based on \cite{Wang2019} and supplementary information:
\begin{equation}
G = \frac{\pi \sigma d^2}{4 l}
\end{equation}
where $d$ represents the diameter of the filament, $\sigma$ denotes the filament's electical conductivity and $l$ its length. Since the electron mean free path of Ag ($\lambda_{\rm f}$ = \SI{53}{\nano\meter}) is longer than the length of a filament (typically <\SI{5}{\nano\meter}) we assume ballistic electron transport. The electrical conductivity can therefore be expressed as follows \cite{Wang2019}:
\begin{equation}
\sigma = \sigma_0 \frac{1 + p}{1 - p} \frac{d}{\lambda_{\rm f}}
\end{equation}
Here, \( \sigma_0 \) is the bulk conductivity, \( \lambda_{\rm f} \) is the electron mean free path in the bulk material (Ag), and \( p \) is the scattering fraction, see Table \ref{tab:values_filament_conductance}.

\begin{table}[h]
    \centering
    \begin{tabular}{c|c|c}
    \hline
         Variable & Value  & Reference \\
         \hline
         $\sigma_0$ & \SI{6.30e7}{\siemens\per\meter} & \cite{Matula1979}\\
         $\lambda_{\rm f}$ &  \SI{53}{\nano\meter} & \cite{Gall2016} \\
         $p$ &  0.5 & \cite{Wang2019} Supplementary Note 9.\\
         \hline
    \end{tabular}
    \caption{Parameters for calculating the conductance of a filament.}
    \label{tab:values_filament_conductance}
\end{table}

\subsubsection{Tunneling Conductance}\label{sec:tunnel_conductance}

\textbf{Simmons Tunneling Theory}\\
The model used for the tunneling conductivity is based on Simmons tunneling theory \cite{Simmons1963a, Zhang2015}, reason being the metal-insulator-metal structure the gaps have. Simmons tunneling formula for general $V_{g}$ brings

\begin{eqnarray}
    J &=& \frac{6.2 \times 10^{10}}{\Delta x^2} \left( \varphi_{I} e^{-1.025\Delta x \sqrt{\varphi_{I}}} - \left( \varphi_{I} + V_{g} \right) e^{-1.025\Delta x \sqrt{ \varphi_{I} + V_{g} }} \right) \nonumber \\ && \times \left( 1 + \frac{3 \times 10^{-9} \Delta x^2 T^2}{\varphi_{I}}\right) \\ 
    \varphi_{I} &=& \varphi_{0} - \left( \frac{V_{g}}{2D} \right) \left( x_{1} + x_{2} \right) - \left( \frac{5.75}{2 \epsilon_{r} \Delta x} \right) \ln \left( \frac{ x_{2} \left( D - x_{1} \right) }{ x_{1} \left( D - x_{2} \right) } \right) \\
    \Delta x &=& x_{2} - x_{1} \\
    x_{1} &=& \frac{3}{\epsilon_{r} \varphi_{0}} \\
    x_{2} &=& D \left( 1 - \frac{46}{6 \varphi_{0} \epsilon_{r} D + 20 - 4 V_{g} \epsilon_{r} D} \right) + x_{1} \quad , V_{g} < \varphi_{0} \\
    x_{2} &=& \frac{ \varphi_{0} \epsilon_{r} D - 14 }{ \epsilon_{r} V_{g} } \hspace{4.4cm} , V_{g} > \varphi_{0} \\
    \varphi_{0} &=& \frac{ W - X }{e}
\end{eqnarray}
where $J$ is the tunneling current density in A/cm$^2$, $D$ is the tunneling distance in \AA, $\epsilon_{r}$ is the relative gap permittivity, $T$ is the temperature in K, $W$ is the work function of the metal and $X$ is the work function of the insulator.

\textbf{Parameters for the Ag NPN Simulation}\\
The tunneling gaps of the network consist of two Ag NPs with vacuum in between. This leads to $W = 4.26$eV, $X = 0$eV, $T = 300$K and $\epsilon_{r} = 1$. With these parameters there occurs a discontinuity at $V_{g} = \varphi_{0}$ that jumps from a greater value to a smaller one. This has the effect of a negative conductivity and can lead to multiple valid solutions when solving for the particles potentials and makes solving algorithms more difficult. To oppose this the weighted mean of $J$ is taken for a small interval around $V_{g} = \varphi_{0}$ to smoothly transition between the curves.

\begin{eqnarray}
    J &=& \frac{\varphi_{0} - V_{g} + \varphi_{w}}{2\varphi_{w}} J_{\text{lower}} + \frac{ V_{g} - \varphi_{0} + \varphi_{w}}{2\varphi_{w}} J_{\text{upper}}
\end{eqnarray}

where $2\varphi_{w}$ in eV is the interpolation width. For the simulation, $2\varphi_{w}$ is set to be $0.4$ eV. $J_{\text{lower}}$ is the tunneling current density for $V_{g} < \varphi_{0}$ and $J_{\text{upper}}$ for $V_{g} > \varphi_{0}$ respectively.\\

\textbf{Linearized Tunneling Model}\\
The non-linearities in the J-V dependence for the tunnel gaps lead to a
nonlinear system of equations in the solution of the particle potentials. This is computationally very expensive to solve, so a model with a linear J-V-dependency is established as proposed in literature \cite{pike_2020},
\begin{eqnarray}
    J = a V_{g} \, e^{-bD}.
\end{eqnarray}
where $a$ in S/cm$^2$ and $b$ in 1/m are model parameters. Since the Simmons formula exhibits an exponential current density-voltage (J-V) dependency, the error function $E$ is chosen as the sum of the squared relative errors to optimize the parameters.
\begin{eqnarray}
    E &=& \frac{1}{N} \sum_{n=1}^{N} \left( \frac{J_{\text{linear},n} - J_{\text{Simmons},n}}{J_{\text{Simmons},n}} \right)^2
\end{eqnarray}

This minimizes the discrepancy between the data points and the model. The fit uses $N=10^8$ data points that are equally spaced on the domain $\left[0.01\text{V}, 10\text{V}\right] \times \left[10\text{\AA},40\text{\AA}\right]$. The resulting parameter values are $a = 1.434 \times 10^{13}$ S/cm$^2$ and $b = 2.094$ 1/\AA.

\textbf{Effective Tunneling Area}\\
A noticable difference from Simmons tunneling theory is the geometry, where for Simmons the tunneling occurs between plates and in the simulation the tunneling is between NPs. A first estimate for the effective tunneling area is evaluated to be $10\%$ of the mean area of the circles generated by the particles radii.

\begin{eqnarray}
    A_{\text{eff}} &=& 0.1 \frac{\pi \left( r_{1}^2 + r_{2}^2 \right) }{2}
\end{eqnarray}

\textbf{Gap Conductance}\\
The gap conductance $G$ is then calculated via
\begin{eqnarray}
    G &=& A_{\text{eff}} \text{ } a e^{-bD}
\end{eqnarray}

\section{Simulation}\label{sec:simlation}
A simulation consists of the initialization: a number of NPs $N$\textsubscript{p} is determined from Eq. \ref{eq:coverage} and deposited on the substrate. Following this, the groups and connections are identified, and each connection is assigned a conductance value. In the main loop of the simulation, the nodal potential analysis is applied over a specified number of iterations - using each NP, or the center of each disk, as a node (Fig. \ref{fig:potential}) - followed by the execution of the gap switching model. This process provides the updated conductance values.

In the simulations carried out here, the number of NPs on the substrate is determined in order to achieve a coverage of $p=0.65$, i.e. slightly below the percolation threshold of $\approx 0.68$ \cite{Mertens2012}. We assume a NP size distribution as shown in Fig. \ref{fig:histogram}.

To analyze the relationship between the switching dynamics and the potential profile, we limit ourselves to a substrate size of 500 nm x 500 nm. Fig. \ref{fig:particles_small} shows the distribution of gaps in different switching states. In Fig. \ref{fig:potential_small}, the potential corresponding to these snapshots is plotted. It is clearly visible that closing the gaps, i.e., complete filament formation between the groups, brings both groups to the same potential. (Re)opened gaps, i.e., tunneling connections between the groups, cause a comparatively high voltage drop between these groups.




\section{Conclusion and Outlook}
We have demonstrated how the current state of modeling and simulation of NPNs can be extended in a physics-based manner. With this new modeling platform, we have created the possibility to study the influences of polydispersity and material properties of the NPs on the switching behavior. In future work, we will investigate the criticality of the NPNs and also study the relationship between polydispersity and material properties on this criticality.

\backmatter



\bmhead{Acknowledgements}

This work was funded by the Deutsche
Forschungsgemeinschaft (DFG, German Research Foundation) - Project-ID 434434223 - SFB 1461.

\section*{Author contributions}

TH: Conceptualization, Modeling, Software, Writing-original draft.
RS: Modeling, Software, Writing-original draft.
SY: Conceptualization, Writing-review and editing.
TG: Conceptualization, Writing-review and editing.
JT: Funding, Writing-review and editing, Supervision
TM: Funding, Writing-review and editing, Supervision

\bmhead{Data Availability Statement} Data will be made available on reasonable request.

\bmhead{Software} The model described here is implemented in Python, utilizing the libraries NumPy, SciPy, and Matplotlib.

\bmhead{Declarations}

\bmhead{Conflict of interest} The authors have no relevant financial or non-financial interests to disclose.

\noindent

\bigskip

\clearpage

\begin{appendices}

\section{Deposition of NP as Poisson Process}\label{secA1}
\begin{eqnarray}
    P_{A}(k) &=& \lim_{n \to \infty} \binom{n}{k} p^k (1-p)^{n-k} \\
    &=& \lim_{n \to \infty} \frac{n!}{k!(n-k)!}\left(\frac{\lambda A}{n}\right)^k \left(1-\frac{\lambda A}{n}\right)^{n-k} \\
    &=& \lim_{n \to \infty} \frac{\sqrt{2\pi n}\left(\frac{n}{e}\right)^n}{k!\sqrt{2\pi (n-k)}\left(\frac{n-k}{e}\right)^{n-k}} \left(\frac{\lambda A}{n}\right)^k \left(1-\frac{\lambda A}{n}\right)^{n-k} \\
    &=& \lim_{n \to \infty} \frac{n^k}{k!} \left(\frac{\lambda A}{n}\right)^k \left(1-\frac{\lambda A}{n}\right)^{n-k} \\
    &=& \lim_{n \to \infty} \frac{1}{k!} (\lambda A)^k \left(1-\frac{\lambda A}{n}\right)^{n-k} \\
    &=& \frac{(\lambda A)^k e^{-\lambda A}}{k!}
\end{eqnarray}




\end{appendices}


\clearpage

\bibliography{sn-bibliography}

\clearpage

\setcounter{figure}{0}
\renewcommand{\thefigure}{\arabic{figure}}

\begin{figure}[t]
\centering
\includegraphics[width=1\textwidth]{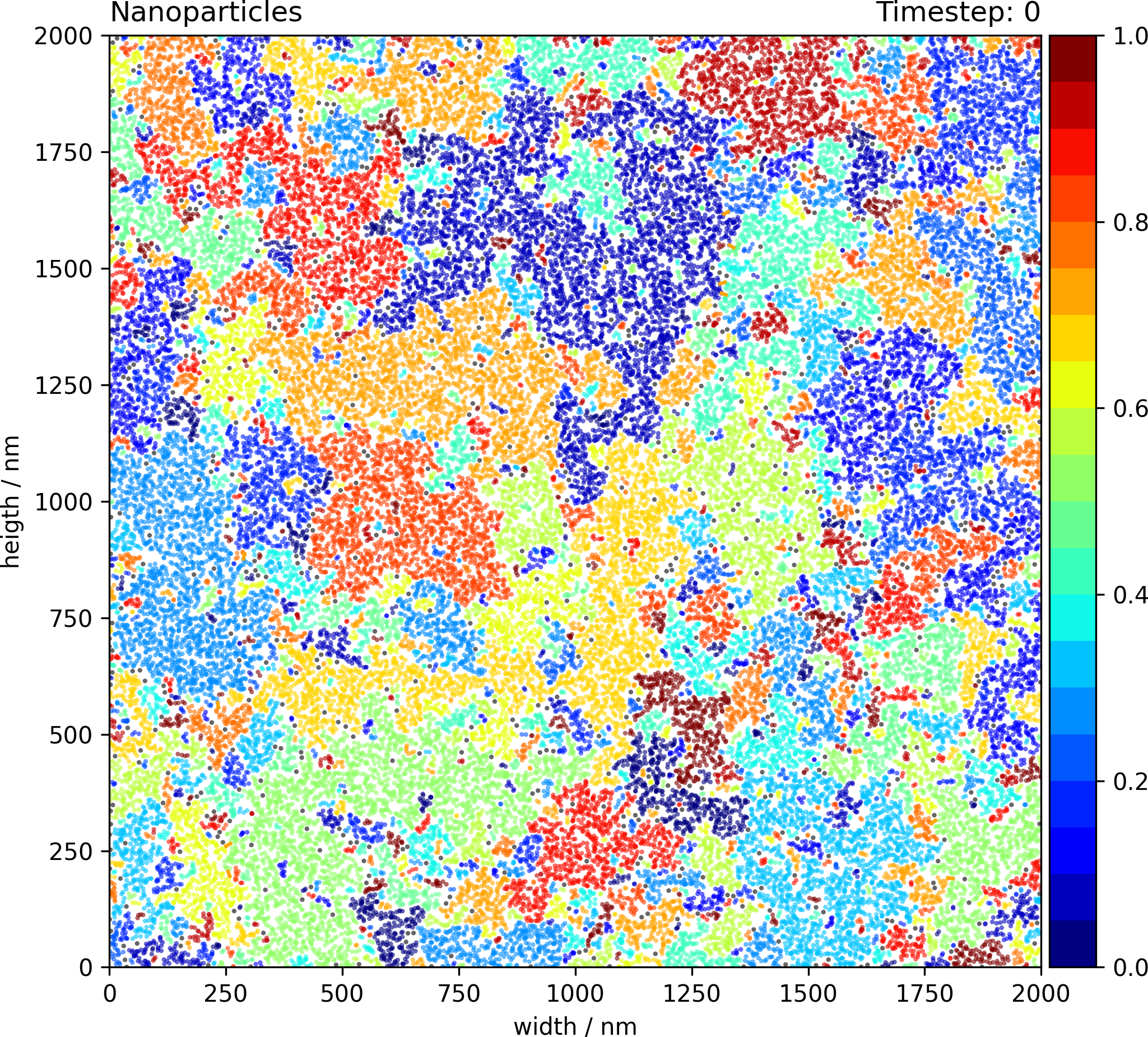}
\caption{2D substrate with NPs. NPs of the same group have the same color, whereby the color of the group is chosen at random.}\label{fig:particles}
\end{figure}

\clearpage

\begin{figure}[t]
    \centering
    \includegraphics[width=1\textwidth]{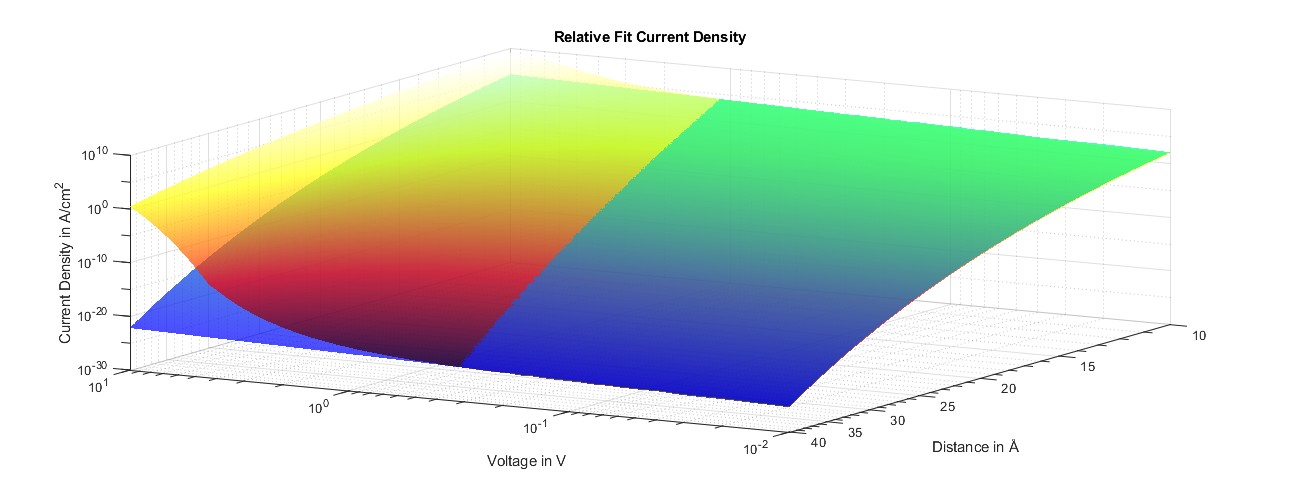}
    \caption{Current density using Simmons model (hot color-map) vs. the linearized model (winter color-map). They match well for low voltages but split apart for higher voltages, especially when the gap length is large.}
    \label{fig:enter-label}
\end{figure}

\clearpage

\begin{figure}[t]
\centering
\includegraphics[width=1\textwidth]{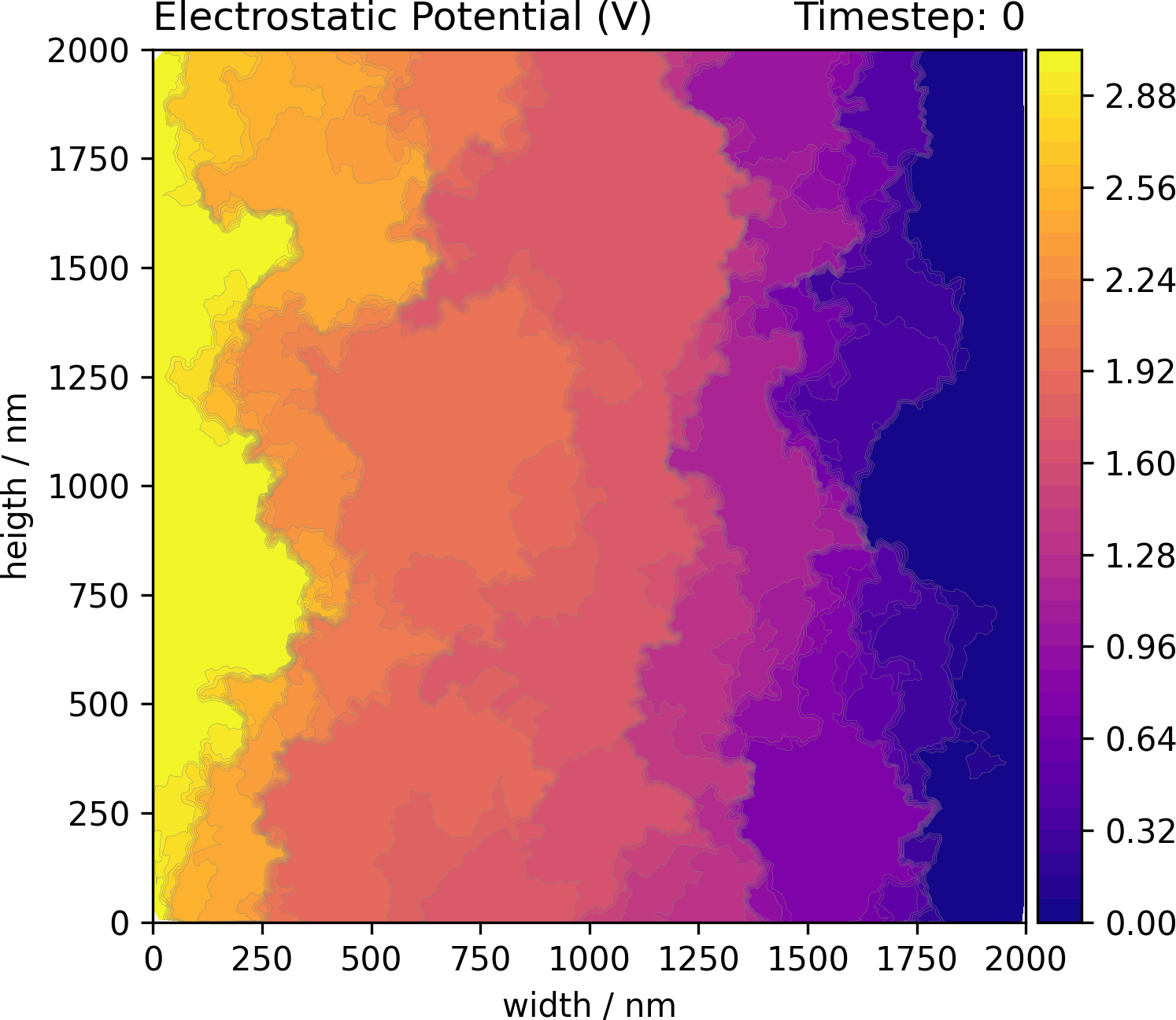}
\caption{Potential map (V) for the initial state before the switching dynamics happen.}\label{fig:potential}
\end{figure}

\clearpage

\begin{figure}[t]
\centering
\includegraphics[width=1\textwidth]{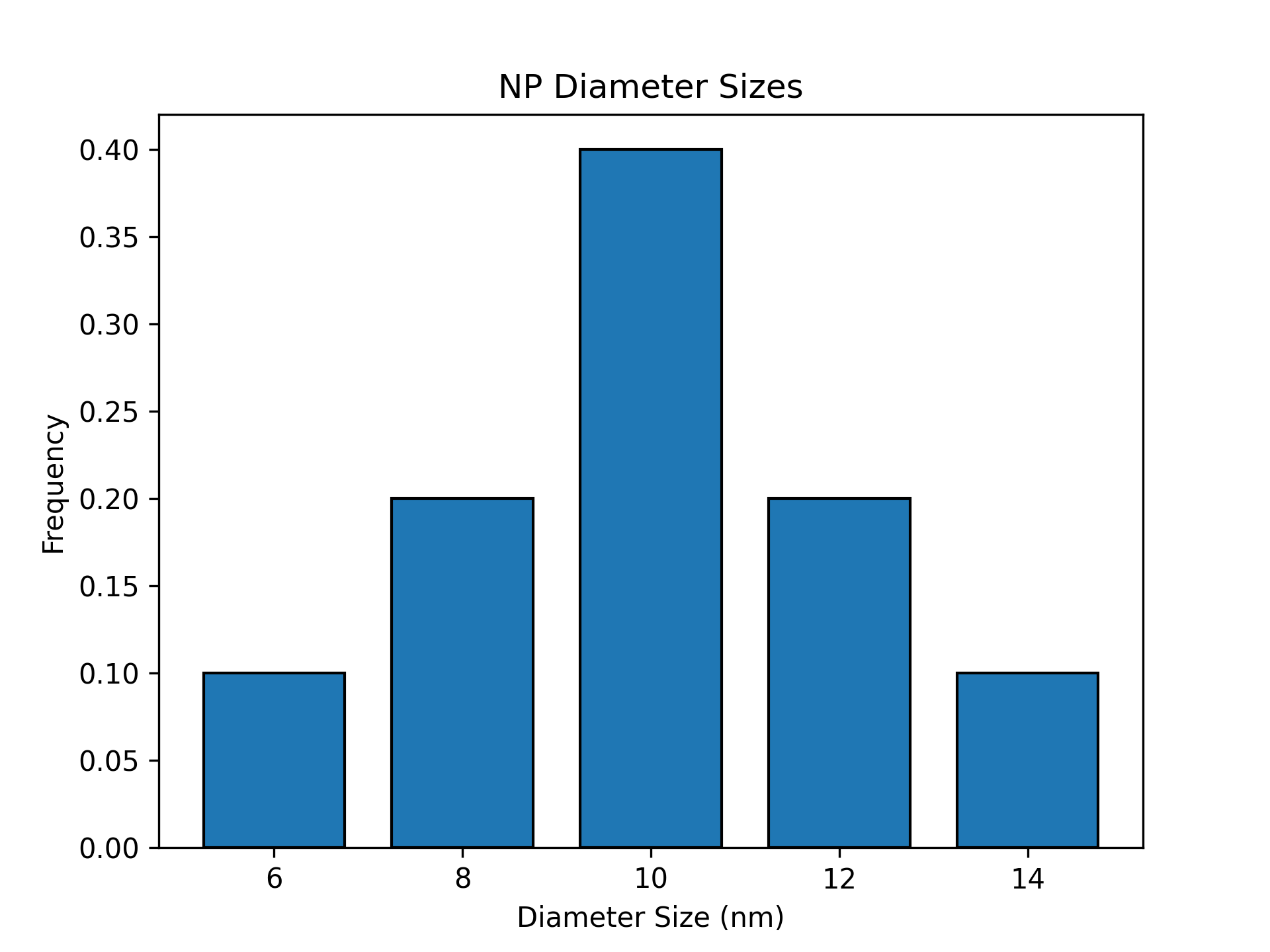}
\caption{Histogram of NP diameters.}\label{fig:histogram}
\end{figure}

\clearpage

\begin{figure}[t]
\centering
\includegraphics[width=0.9\textwidth]{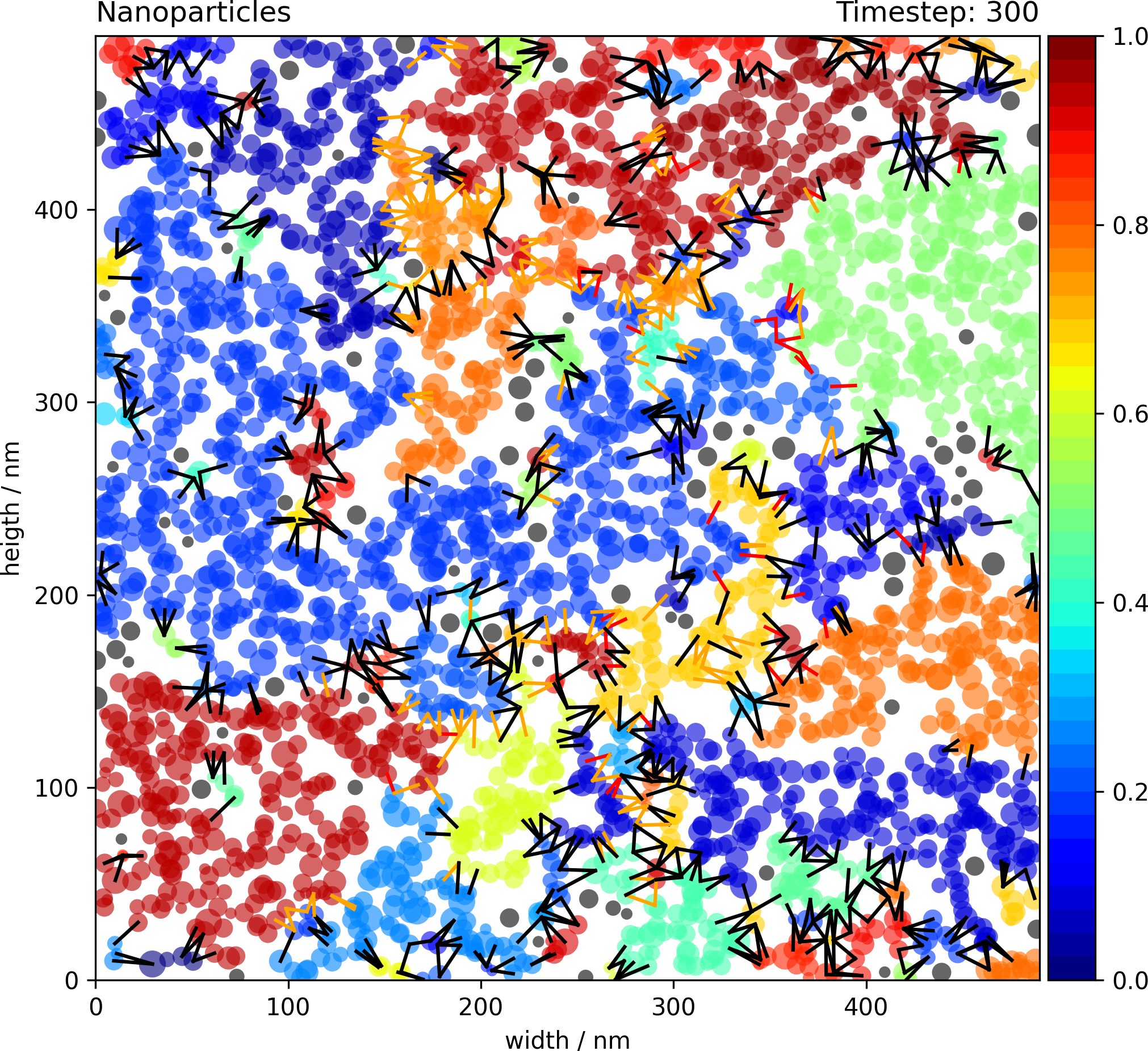}\vspace{1cm}
\includegraphics[width=0.9\textwidth]{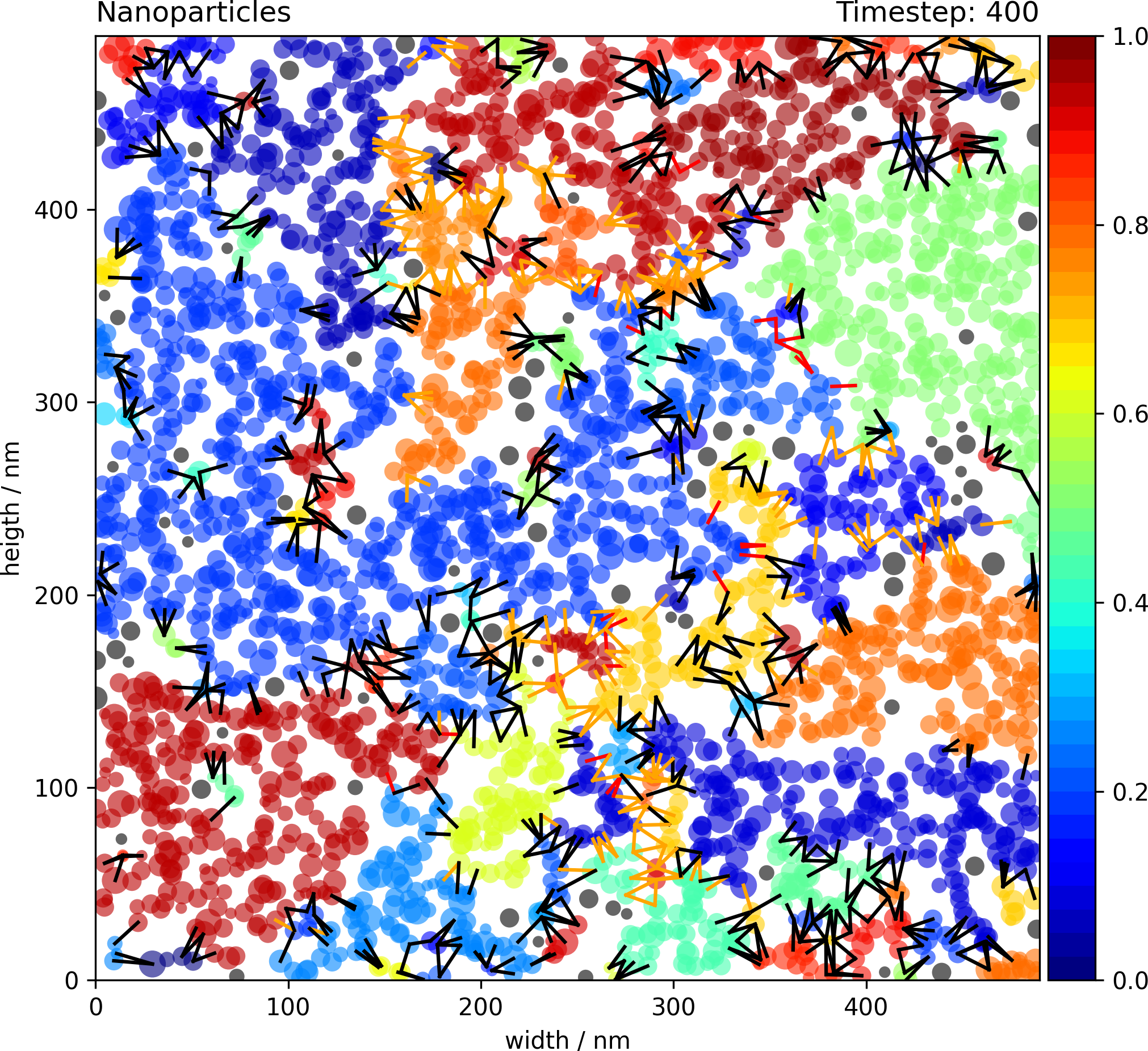}
\caption{NPN with switching gaps at different time steps: 
Tunnel gaps are drawn in black, closing gaps (growing filaments) in orange, and closed gaps in red.}\label{fig:particles_small}
\end{figure}

\clearpage

\begin{figure}[t]
\centering
\includegraphics[width=0.9\textwidth]{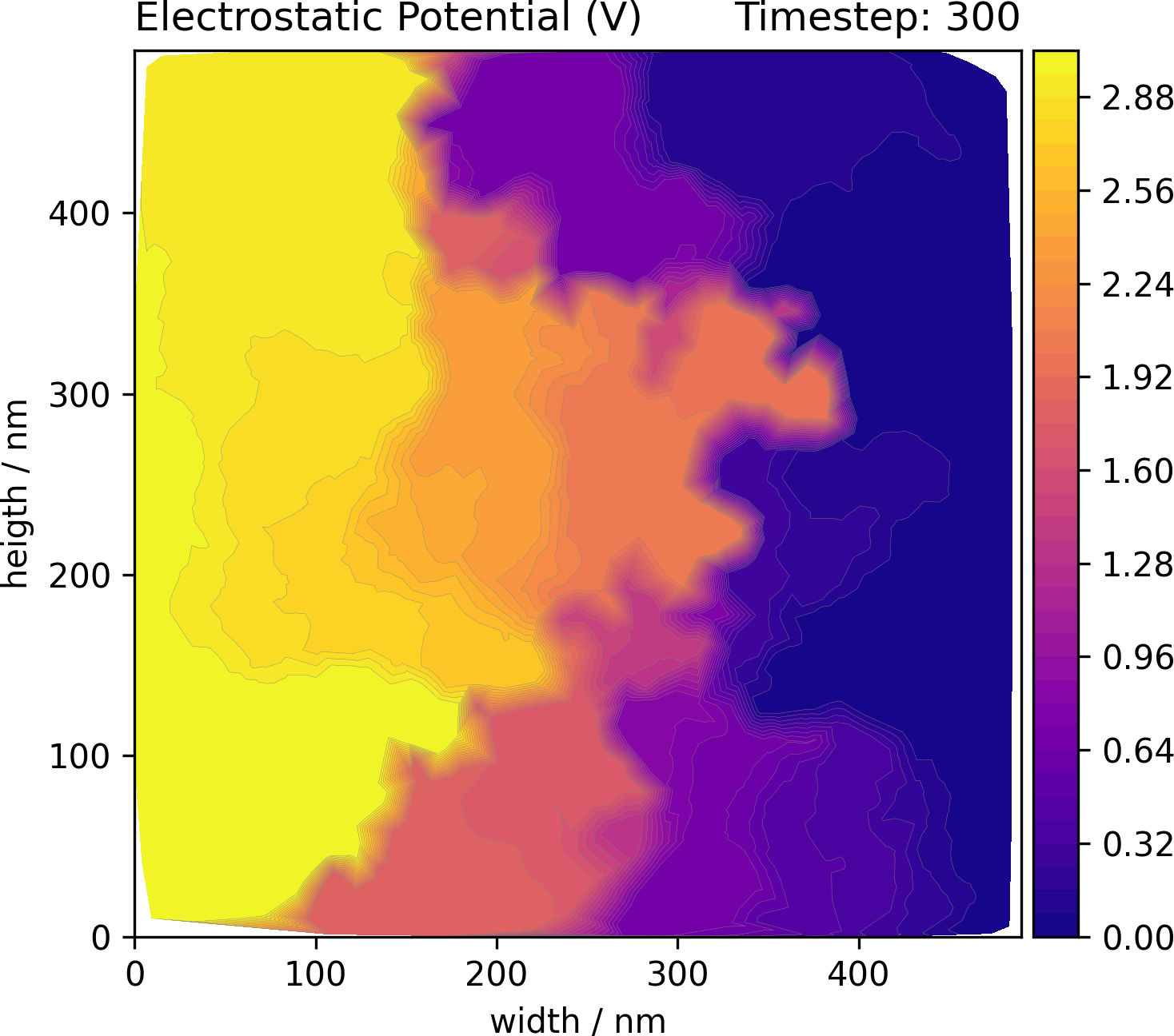}
\includegraphics[width=0.9\textwidth]{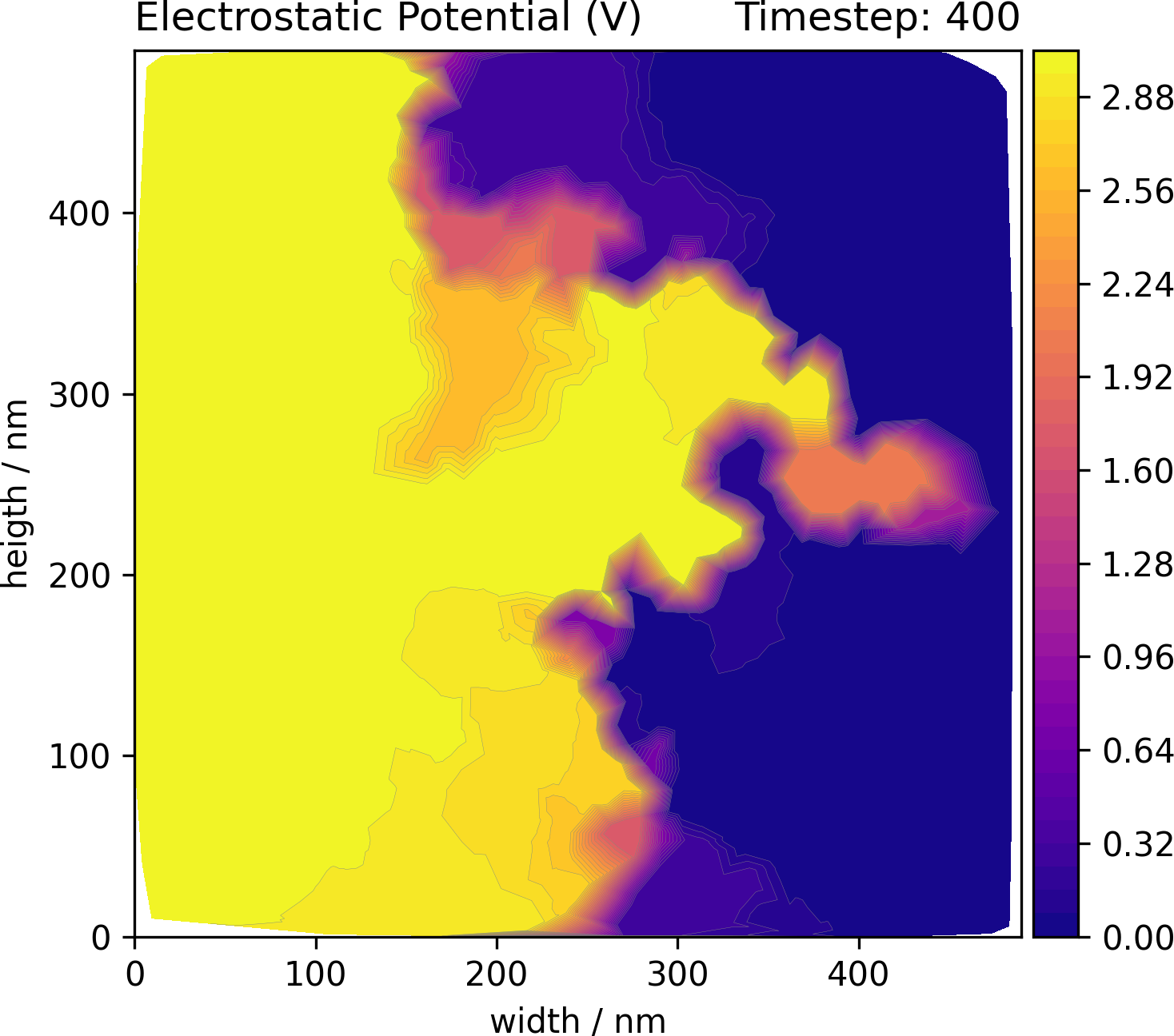}
\caption{Potential maps at corresponding time steps, of Fig. \ref{fig:particles_small}}\label{fig:potential_small}
\end{figure}

\clearpage

\begin{figure}[t]
\centering
\includegraphics[width=0.8\textwidth]{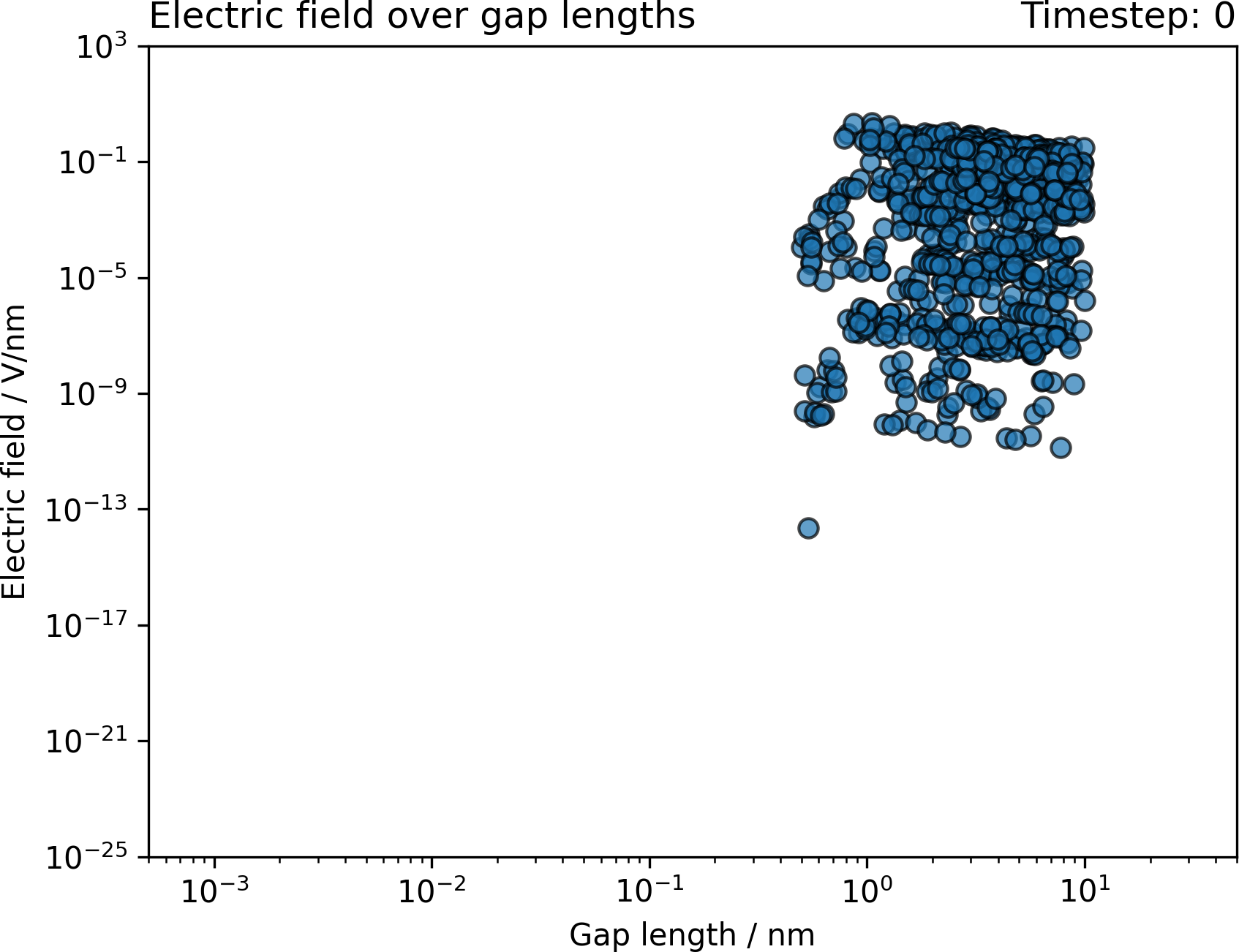}
\includegraphics[width=0.8\textwidth]{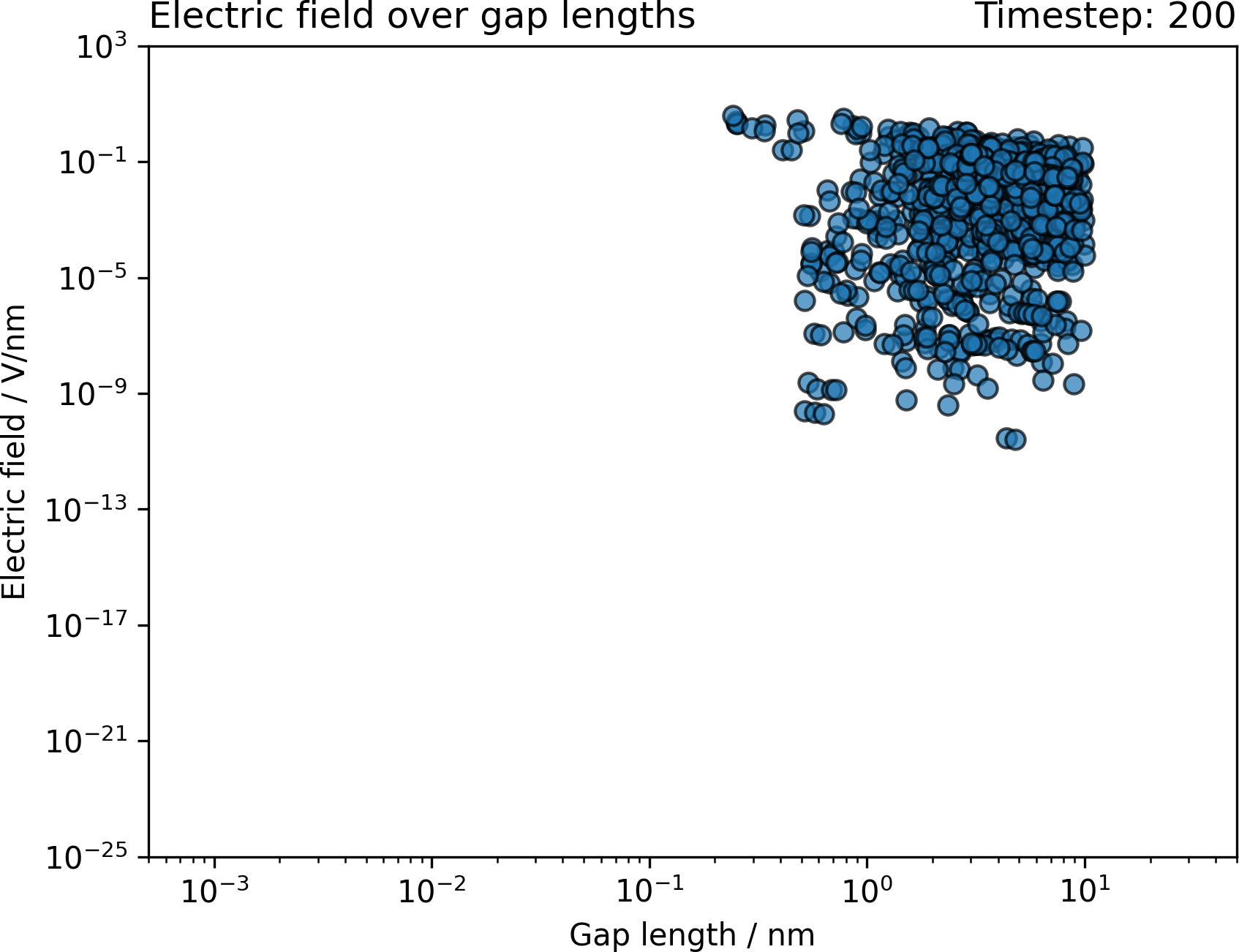}
\caption{Electric fields in the gaps plotted over the gap length at different time steps.}\label{fig:efield-gap-length}
\end{figure}

\end{document}